\documentclass[sigconf]{acmart}

\usepackage{array}
\usepackage{amsmath}
\usepackage{amsfonts}
\usepackage{amssymb}
\usepackage{multirow}
\usepackage{tabularx}
\usepackage{graphicx}
\usepackage[para,online,flushleft]{threeparttable}

\def\BibTeX{{\rm B\kern-.05em{\sc i\kern-.025em b}\kern-.08emT\kern-.1667em\lower.7ex\hbox{E}\kern-.125emX}}
    
\copyrightyear{2019} 
\acmYear{2019} 
\setcopyright{rightsretained} 
\acmConference[KDD '19]{The 25th ACM SIGKDD Conference on Knowledge Discovery and Data Mining}{August 4--8, 2019}{Anchorage, AK, USA}
\acmBooktitle{The 25th ACM SIGKDD Conference on Knowledge Discovery and Data Mining (KDD '19), August 4--8, 2019, Anchorage, AK, USA}
\acmDOI{10.1145/3292500.3330759}
\acmISBN{978-1-4503-6201-6/19/08}

\input{Definitions}

\begin{document}

%
\title{Semantic Product Search}

%

\author{Priyanka Nigam}
\authornote{Both authors contributed equally to this research.}
\email{nigamp@amazon.com}
\affiliation{%
	\institution{Amazon}
	\city{Palo Alto}
	\state{California}
	\country{USA}
}

\author{Yiwei Song}
\authornotemark[1]
\email{ywsong@amazon.com}
\affiliation{%
	\institution{Amazon}
	\city{Palo Alto}
	\state{California}
	\country{USA}
}

\author{Vijai Mohan}
\email{vijaim@amazon.com}
\affiliation{%
	\institution{Amazon}
	\city{Palo Alto}
	\state{California}
	\country{USA}
}

\author{Vihan Lakshman}
\email{vihan@amazon.com}
\affiliation{%
	\institution{Amazon}
	\city{Palo Alto}
	\state{California}
	\country{USA}
}

\author{Weitian (Allen) Ding}

\authornote{This work was done while the author was at Amazon}
\email{weitiand@andrew.cmu.edu}
\affiliation{%
	\institution{Carnegie Mellon University}
	\city{Pittsburgh}
	\state{Pennsylvania}
	\country{USA}
}

\author{Ankit Shingavi}
\email{ashingav@amazon.com}
\affiliation{%
	\institution{Amazon}
	\city{Palo Alto}
	\state{California}
	\country{USA}
}

\author{Choon Hui Teo}
\email{choonhui@amazon.com}
\affiliation{%
	\institution{Amazon}
	\city{Palo Alto}
	\state{California}
	\country{USA}
}

\author{Hao Gu}
\email{hggu@amazon.com}
\affiliation{%
	\institution{Amazon}
	\city{Palo Alto}
	\state{California}
	\country{USA}
}

\author{Bing Yin}
\email{alexbyin@amazon.com}
\affiliation{%
	\institution{Amazon}
	\city{Palo Alto}
	\state{California}
	\country{USA}
}

%
\renewcommand{\shortauthors}{Nigam and Song et al.}

%
\begin{abstract}
  We study the problem of semantic matching in product search, that is,
  given a customer query, retrieve all semantically related products from
  the catalog. Pure lexical matching via an inverted index falls short
  in this respect due to several factors: a) lack of understanding of
  hypernyms, synonyms, and antonyms, b) fragility to morphological
  variants (e.g. ``woman" vs. ``women"), and c) sensitivity to spelling
  errors. To address these issues, we train a deep learning model for
  semantic matching using customer behavior data. Much of the recent work on
  large-scale semantic search using deep learning focuses on ranking for
  web search. In contrast, semantic matching for product search presents
  several novel challenges, which we elucidate in this paper. We
  address these challenges by a) developing a new loss function that has an inbuilt
  threshold to differentiate between random negative examples, impressed but not purchased
  examples, and positive examples (purchased items), b) using average pooling in conjunction
  with $n$-grams to capture short-range linguistic patterns, c) using
  hashing to handle out of vocabulary tokens, and d) using
  a model parallel training architecture to scale
  across 8 GPUs. We present compelling offline results that
  demonstrate at least 4.7\% improvement in Recall@100 and 
  14.5\% improvement in mean average precision (MAP) 
  over baseline state-of-the-art semantic search methods using the same tokenization method. 
  Moreover, we present results and discuss learnings from online A/B
  tests which demonstrate the efficacy of our method.
\end{abstract}

%
%

\ccsdesc[500]{Information systems~Retrieval models and ranking}
\ccsdesc[500]{Applied computing~Electronic commerce}

%
\keywords{Semantic Matching, Product Search, Neural Information Retrieval}

%
\maketitle
\section{Introduction}

\begin{figure}[h]
	\centering
	\includegraphics[height=4cm, width=\linewidth]{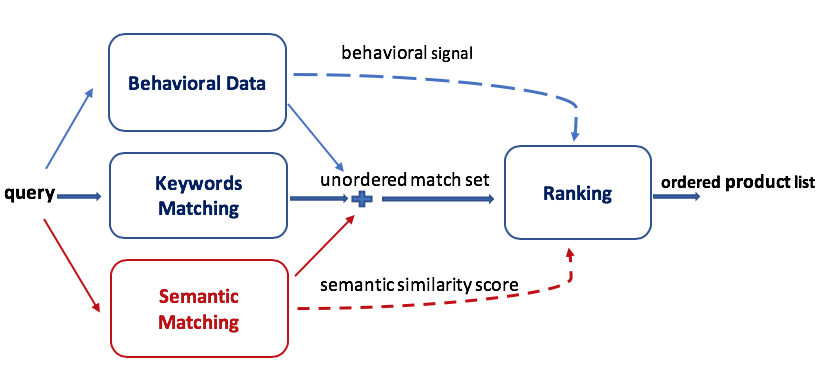}
	\caption{System architecture for augmenting product matching using semantic matching }
	\label{fig:system_architecture}
\end{figure}

At a high level, as shown in Figure \ref{fig:system_architecture}, a product search engine works as follows: a
customer issues a query, which is passed to a lexical matching engine
(typically an inverted index \cite{zobel2006inverted,manning2008introduction}) to retrieve all
products that contain words in the query, producing a \emph{match set}. The
match set passes through stages of ranking, wherein top
results from the previous stage are re-ranked before
the most relevant items are finally displayed. It is imperative that the match
set contain a relevant and diverse set of products that match the customer
intent in order for the subsequent rankers to succeed. However,
inverted index-based lexical matching falls short in several key
aspects:
\begin{itemize}
\item Lack of understanding of hypernyms (generalizations of words),
  synonyms (different words with the same meaning), and antonyms (words
  that have opposite meanings). For example, ``sneakers" might match the intent of the query
  \emph{running shoes}, but may not be retrieved. Similarly, a ``red
  dress" matches the semantic intent of the query \emph{burgundy dress}
  and yet is not retrieved by a lexical matching engine. Finally, ``latex
  free examination gloves'' do not match the intent of the query
  \emph{latex examination gloves}, and yet are retrieved simply because
  all the words in the query are also present in the product title and
  description.
\item Fragility to morphological variants (e.g. ``woman" vs. \\``women"). One can
  address this issue to some extent by applications of stemming or lemmatization. However,
  stemming algorithms are often imperfect and lead to information loss
  and errors. For instance, a stemmer that truncates nouns into their singular form might transform
  the query ``reading glasses" into ``reading glass" and fail to return relevant results. To be viable in production,
  these approaches typically require numerous hand-crafted rules that may become obsolete and fail to
  generalize to multiple languages. 
\item Sensitivity to spelling errors. According to some estimates of web search logs \cite{cucerzan2004spelling,dalianis2002evaluating}, 10-15\% of queries are misspelled. This
  leads to customer confusion (why are there no results for the query
  ``rred drress''?) and frustration. While modern spell-correction
  methods can reduce the problem, a matching engine that handles spelling errors would be simpler. 
\end{itemize}
In this paper, we address the question: Given rich
customer behavior data, can we train a deep learning model to retrieve matching products in response to a query? Intuitively,
there is reason to believe that customer behavior logs
contain semantic information; customers who are intent on purchasing a
product circumvent the limitations of lexical matching by
query reformulation or by deeper exploration of the search
results. The challenge is the sheer magnitude of the data as
well as the presence of noise, a challenge that modern deep learning
techniques address very effectively. 

Product search is different from web search as the queries tend to be shorter and the positive signals (purchases) are sparser than clicks. Models based on conversion rates or click-through-rates may incorrectly favor accessories (like a phone cover)
over the main product (like a cell phone). This is further complicated
by shoppers maintaining multiple intents during a single search
session: a customer may be looking for a specific television model
while also looking for accessories for this item at the lowest
price and browsing additional products to qualify for free
shipping.  A product search engine should reduce the effort needed from a customer with a specific mission (narrow queries) while allowing shoppers to explore when they are looking for inspiration (broad queries). 

As mentioned, product search typically operates in two stages: matching and ranking. Products that contain words in the query ($Q_i$) are the primary candidates. Products that have prior behavioral associations (products bought or clicked after issuing a query $Q_i$) are also included in the candidate set. The ranking step takes these candidates and orders them using a machine-learned rank function to optimize for customer satisfaction and business metrics.

We present a neural network trained with large amounts of purchase and click signals to complement a lexical search engine in ad hoc product retrieval. Our first contribution is a loss function with a built-in threshold to differentiate between random negative, impressed but not purchased, and purchased items. Our second contribution is the empirical result that recommends average pooling in combination with $n$-grams that capture short-range linguistic patterns instead of more complex architectures. Third, we show the effectiveness of consistent token hashing in Siamese networks for zero-shot learning and handling out of vocabulary tokens. 

In Section \ref{sec:relatedwork}, we highlight related work. In Section \ref{sec:model}, we describe our model architecture, loss functions, and tokenization techniques including our approach for unseen words. We then introduce the readers to the data and our input representations for queries and products in Section \ref{sec:data}. Section \ref{sec:expt} presents the evaluation metrics and our results. We provide implementation details and optimizations to efficiently train the model with large amounts of data in Section \ref{sec:train_acc}. Finally, we conclude in Section \ref{sec:conclusion} with a discussion of future work. 

\section{Related Work}
\label{sec:relatedwork}

There is a rich literature in natural language processing (NLP)
and information retrieval (IR) on capturing the semantics of queries
and documents. Word2vec \cite{mikolov2013distributed} garnered significant
attention by demonstrating the use of word embeddings to capture
semantic structure; synonyms cluster together in the embedding
space. This technique was successfully applied to document ranking for web search with the DESM model \cite{mitra2016dual}. Building from the ideas in word2vec, \citet{diaz2016query} trained neural word embeddings to find neighboring words to expand queries with synonyms. Ultimately, based on these recent advancements and other key insights, the state-of-the-art models for semantic search can generally be classified into three categories:
 
\begin{enumerate}
\item  {\verb|Latent Factor Models|}: Nonlinear matrix completion approaches that learn query and document-level embeddings without using their content.
\item  {\verb|Factorized Models|}: Separately convert queries and documents to low-dimensional embeddings based on content.
\item  {\verb|Interaction Models|}: Build interaction matrices between the query and document text and use neural networks to mine patterns from the interaction matrix
\end{enumerate}

\citet{deerwester1990indexing} introduced Latent Semantic Analysis (LSA), which computes a low-rank factorization of a term-document matrix to identify semantic concepts and was further refined by \cite{berry1995using, dumais1997automatic} and extended by ideas from Latent Dirichlet Allocation (LDA) \cite{blei2003latent} in \cite{wei2006lda}.
 In 2013, \citet{huang2013dssm} published the seminal paper in the space of factorized models by introducing the Deep Semantic Similarity Model (DSSM). Inspired by LSA and Semantic Hashing \cite{salakhutdinov2009semantic}, DSSM involves training an end-to-end deep neural network with a discriminative loss to learn a fixed-width representation for queries and documents. Fully connected units in the DSSM architecture were subsequently replaced with Convolutional Neural Networks (CNNs) \cite{shen2014latent, Hu:2014:CNN:2969033.2969055} and Recurrent Neural Networks (RNNs) \cite{palangi2016deep} to respect word ordering. In an alternate approach, which articulated the idea of interaction models, \citet{guo2016deep} introduced the Deep Relevance Matching Model (DRMM) which leverages an \emph{interaction matrix} to capture local term matching within neural approaches which has been successfully extended by MatchPyramid \cite{pang2016text} and other techniques \cite{yang2016anmm, wan2016match, hui2017pacrr, hui2017re, hui2018co}. Nevertheless, these interaction methods require memory and computation proportional to the number of words in the document and hence are prohibitively expensive for online inference. In addition, Duet \cite{mitra2017duet} combines the approaches of DSSM and DRMM to balance the importance of semantic and lexical matching. Despite obtaining state-of-the-art results for ranking, these methods report limited success on ad hoc retrieval tasks \cite{mitra2017duet} and only achieve a sub-50\% Recall@100 and MAP on our product matching dataset, as shown with the ARC-II and Match Pyramid baselines in Table \ref{tab:baselines}.

While we frequently evaluate our hypotheses on interaction matrix-based methods, we find that a factorized model architecture achieves comparable performance while only requiring constant memory per product. Hence, we only present our experiments as it pertains to factorized models in this paper. Although latent factor models improve ranking metrics due to their ability to memorize associations between the query and the product, we exclude it from this paper as we focus on the matching task. Our choice of model architecture was informed by empirical experiments while constrained by the cost per query and our ability to respond within 20 milliseconds for thousands of queries per second.

\section{Model}
\label{sec:model}

\subsection{Neural Network Architecture}
\label{sec:arch}
\begin{figure}[h]
	\centering
	\includegraphics[width=\linewidth]{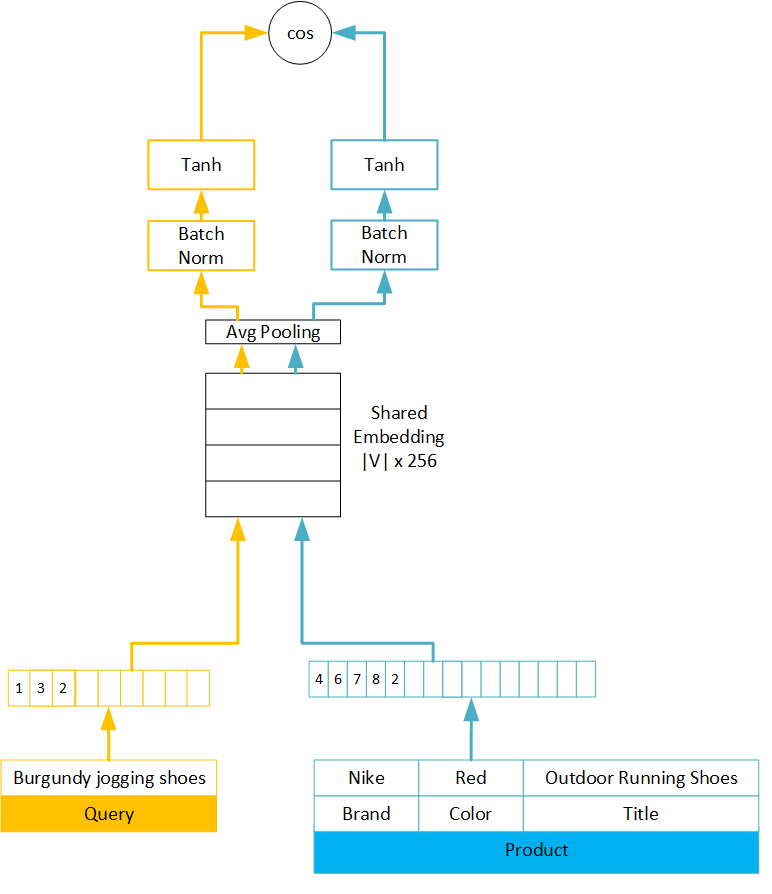}
	\caption{Illustration of neural network architecture used for semantic search}
	\label{fig:neural_architecture}
\end{figure}

Our neural network architecture is shown in Figure \ref{fig:neural_architecture}. As in the distributed arm of the Duet model, our first model component is an embedding layer that consists of $|V| \times N$ parameters where $V$ is the vocabulary and $N$ is the embedding dimension. Each row corresponds to the parameters for a word. Unlike Duet, we share our embeddings across the query and product. Intuitively, sharing the embedding layer in a Siamese network works well, capturing local word-level matching even before training these networks. Our experiments in Table \ref{tab:shared_emb} confirm this intuition. We discuss the specifics of our query and product representation in Section \ref{sec:data}.

To generate a fixed length embedding for the query ($E^Q$) and the product ($E^P$) from individual word embeddings, we use average pooling after observing little difference (<0.5\%) in both MAP and Recall@100
 relative to recurrent approaches like LSTM and GRU (see Table \ref{tab:pool}). Average pooling also requires far less computation, reducing training time and inference latency. We reconciled this departure from state-of-the-art solutions for Question Answering and other NLP tasks through an analysis that showed that, unlike web search, both query and product information tend to be shorter, without long-range dependencies. Additionally, product search queries do not contain stop words and typically require every query word (or its synonym) to be present in the product. 

Queries typically have fewer words than the product content. Because of this, we observed a noticeable difference in the magnitude of query and product embeddings. This was expected as the query and the product models were shared with no additional parameters to account for this variance. Hence, we introduced Batch Normalization layers \cite{ioffe2015batch} after the pooling layers for the query and the product arms. Finally, we compute the cosine similarity between $E^Q$ and $E^P$. During online A/B testing, we precompute $E^P$ for all the products in the catalog and use a $k$-Nearest Neighbors algorithm to retrieve the most similar products to a given query $Q_i$. 

\subsection{Loss Function}
\label{sec:loss_fn}
A critical decision when employing a vector space model is defining a match, especially in product search where there is an important tradeoff between precision and recall. For example, accessories like mounts may also be relevant for the query ``led tv.''

Pruning results based on a threshold is a common practice to identify
the match set. Pointwise loss functions, such as mean squared error
(MSE) or mean absolute error (MAE), require an additional step
post-training to identify the threshold. Pairwise loss functions do not
provide guarantees on the magnitude of scores (only on relative
ordering) and thus do not work well in practice with threshold-based
pruning. Hence, we started with a pointwise 2-part hinge loss function
as shown in Equation \eqref{eq:2partloss} that maximizes the similarity between
the query and a purchased product while minimizing the similarity
between a query and random products. Define
$\hat{y} := \cos\rbr{E^{Q} , E^{P}}$, and let $y = 1$ if product $P$ is
purchased in response to query $Q$, and $y = 0$ otherwise. Furthermore
let $\ell_{+}\rbr{y} := (-\min\rbr{0, y - \epsilon_{+}})^{m}$, and
$\ell_{-}\rbr{y} := \max\rbr{0, y - \epsilon_{-}}^{m}$ for some
predefined thresholds $\epsilon_{+}$ and $\epsilon_{-}$ and
$m \in \cbr{1, 2}$. The two part hinge loss can be defined as
\begin{align}
  \label{eq:2partloss}
  L\rbr{\yhat, y} := y \cdot \ell_{+}\rbr{\yhat} + \rbr{1 - y} \cdot \ell_{-}\rbr{\yhat}  
\end{align}
Intuitively, the loss ensures that when $y = 0$ then $\yhat$ is less
than $\epsilon_{-}$ and when $y = 1$ then $\yhat$ is above
$\epsilon_{+}$. After some empirical tuning on a validation set, we set
$\epsilon_{+} = 0.9$ and $\epsilon_{-} = 0.2$. 
\begin{figure}[htbp]
	\centering
	\includegraphics[width=\linewidth]{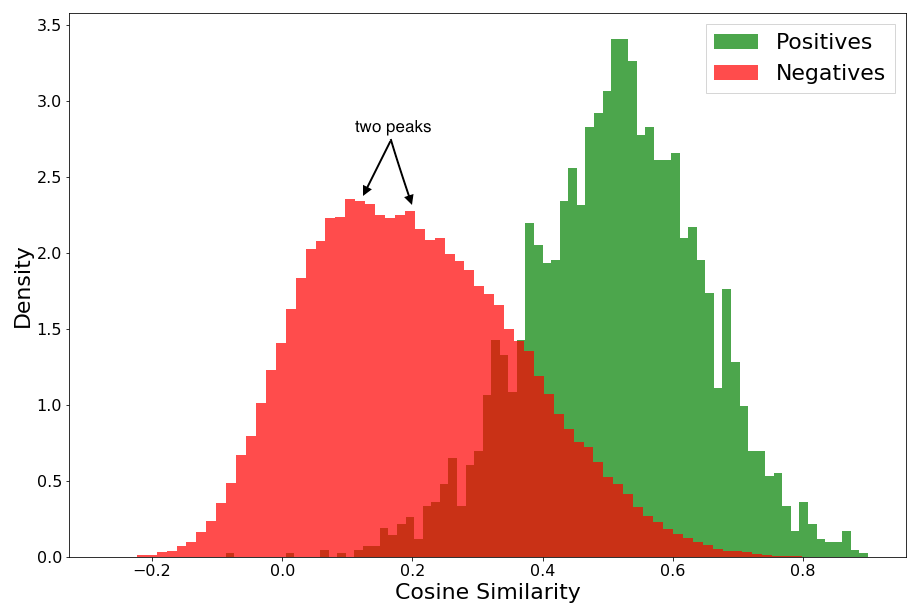}
	\caption{Score distribution histogram shows large overlap for positives (right) and negatives (left) along with a bimodal distribution for negatives when using the 2-part hinge}
	\label{fig:2part}
\end{figure}

\begin{figure}[h]
	\centering
	\includegraphics[width=\linewidth]{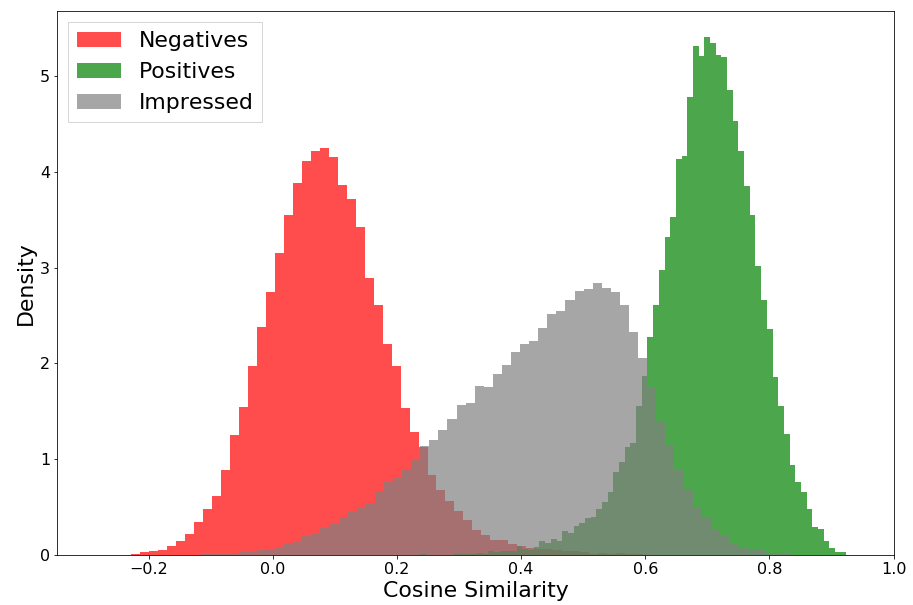}
	\caption{Score distribution shows clear separation between purchased (right), seen but not purchased (center), and irrelevant products (left) when using the 3-part hinge}
	\label{fig:3part}
\end{figure}

As shown in Table \ref{tab:loss}, the 2-part hinge loss improved offline matching 
performance by more than 2X over the MSE baseline. However in
Figure \ref{fig:2part}, a large overlap in score distribution between positives and negatives 
can be seen. Furthermore, the score distribution for 
negatives appeared bimodal. After manually inspecting the negative training examples 
that fell in this region, we uncovered that these were products that were impressed but not 
purchased by the customer. 
From a matching standpoint, these products are usually valid results to show to customers. 
To improve the model's ability to distinguish positives and negatives considering these 
two classes of negatives, we introduced a  3-part hinge loss:
\begin{align}
  \label{eq:3partloss}
  L\rbr{\yhat, y} := I^+(y) \cdot \ell_{+}\rbr{\yhat} + I^-\rbr{y}\cdot \ell_{-}\rbr{\yhat} + I^0\rbr{y} \cdot \ell_{0}\rbr{\yhat}   
\end{align}
where $I^+\rbr{y}$, $I^-\rbr{y}$, and $I^0\rbr{y}$ denote indicators signifying if the product $P$ was purchased, not impressed and not purchased, and impressed (but not purchased) in response to the query $Q$, respectively, and
$\ell_{0}\rbr{\yhat} := \max\rbr{0, \yhat - \epsilon_{0}}^{m}$. Based on the 2-part hinge score distribution, 
$\epsilon_{0}$ was set to $0.55$ with $\epsilon_{+} = 0.9$ and $\epsilon_{-} = 0.2$ as before. The
effectiveness of this strategy can be seen in Figure~\ref{fig:3part},
where one can observe a clear separation in scores between random and
impressed negatives vs positives.

\subsection{Tokenization Methods }
\label{sec:tokenization}

In this section, we describe our tokenization methodology, or the procedure by which we break a string into a sequence of smaller components such as words, phrases, sub-words, or characters. We combine word unigram, word n-gram, and character trigram features into a bag of n-grams and use hashing to handle the large vocabulary size, similar to the fastText approach \cite{joulin2017bag}.

\subsubsection{Word Unigram}
This is the basic form of tokenization where the input query or product title is tokenized into a list of words. For example, the word unigrams of "artistic iphone 6s case" are ["artistic", "iphone", "6s", "case"].

\subsubsection{Word N-gram}
In a bag of words model like ours, word unigrams lose word ordering. Instead of using LSTMs or CNNs to address this issue, we opted for $n$-grams as in \cite{sidorov2014syntactic}. For example, the word bigrams of "artistic iphone 6s case" are ["artistic\#iphone", "iphone\#6s", "6s\#case"] and the trigrams are ["artistic\#iphone\#6s", "iphone\#6s\#case"]. These $n$-grams capture phrase-level information; for example if ``for iphone'' exists in the query, the model can infer that the customer's intention is to search for iphone accessories rather than iphone --- an intent not captured by a unigram model. 

\subsubsection{Character Trigram}
Character trigram embeddings were proposed by the DSSM paper \cite{huang2013dssm}. The string is broken into a list of all three-character sequences. For the example "artistic iphone 6s case", the character trigrams are ["\#ar", "art", "rti", "tis", "ist", "sti", "tic", "ic\#", "c\#i", "\#ip", "iph", "pho", "hon", "one", "ne\#", "e\#6", "\#6s", "6s\#", "s\#c", "\#ca", "cas", "ase", "se\#"]. Character trigrams are robust to typos (``iphione'' and ``iphonr'') and
handle compound words (``amazontv'' and ``firetvstick'') naturally. Another advantage in our setting is the ability to capture similarity of model parts and sizes.

\subsubsection{Handling Unseen Words}

It is computationally infeasible to maintain a vocabulary that includes all the possible word $n$-grams as the dictionary size grows exponentially with $n$. Thus, we maintain a "short" list of several tens or hundreds of thousands of $n$-grams based on token frequency. A common practice for most NLP applications is to mask the input or use the embedding from the $0^{th}$ location when an out-of-vocabulary word is encountered. Unfortunately, in Siamese networks, assigning all unknown words to the same shared embedding location results in incorrectly mapping two different out-of-vocabulary words to the same representation. Hence, we experimented with using the ``hashing trick" \cite{weinberger2009feature} popularized by Vowpal Wabbit to represent higher order $n$-grams that are not present in the vocabulary. In particular, we hash out-of-vocabulary tokens to additional embedding bins. The combination of using a fixed hash function and shared embeddings ensures that unseen tokens that occur in both the query and document map to the same embedding vector. During our initial experiments with a bin size of 10,000, we noticed that hashing collisions incorrectly promoted irrelevant products for queries, led to overfitting, and did not improve offline metrics. However, setting a bin size 5-10 times larger than the vocabulary size improved the recall of the model. 

\subsubsection{Combining Tokenizations}
\begin{figure}[h]
	\centering
	\includegraphics[width=\linewidth]{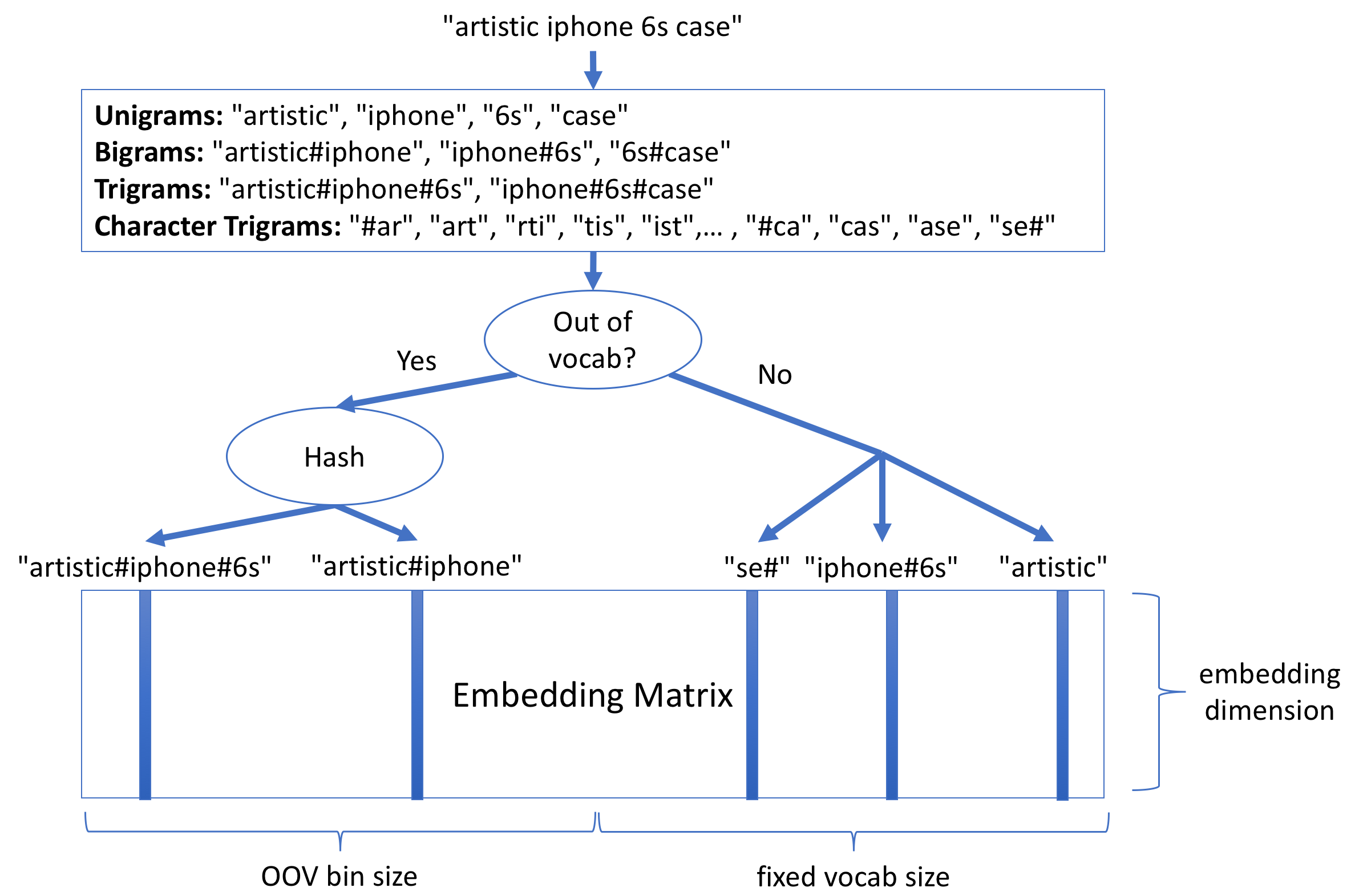}
	\caption{Aggregation of different tokenization methods illustrated with the processing of ``artistic iphone 6s case''}
	\label{fig:aggregated_tokenization}
\end{figure}

There are several ways to combine the tokens from these tokenization methods. One could create separate embeddings for unigrams, bigrams, character trigrams, etc. and compute a weighted sum over the cosine similarity of these $n$-gram projections. But we found that the simple approach of combining all tokens in a single bag-of-tokens performs well. We conclude this section by referring the reader to Figure \ref{fig:aggregated_tokenization}, which walks through our tokenization methods for the example ``artistic iphone 6s case''. In Table \ref{tab:qual}, we show example queries and products retrieved to highlight the efficacy of our best model to understand synonyms, intents, spelling errors and overall robustness. 

\section{Data}
\label{sec:data}

We use 11 months of search logs as training data and 1 month as evaluation. We sample 54 billion query-product training pairs. We preprocess these sampled pairs to 650 million rows by grouping the training data by query-product pairs over the entire time period and using the aggregated counts as weights for the pairs. We also decrease the training time by 3X by preprocessing the training data into tokens and using \texttt{mmap} to store the tokens. More details on our best practices for reducing training time can be found in Section \ref{sec:train_acc}. 

For a given customer query, each product is in exactly one of three categories: purchased, impressed but not purchased, or random. For each query, we target a ratio of 6 impressed and 7 random products for every query-product purchase. We sample this way to train the model for both matching and ranking, although in this paper we focus on matching. Intuitively, matching should differentiate purchased and impressed products from random ones; ranking should differentiate purchased products from impressed ones.

We choose the most frequent words to build our vocabulary, referred to as $|V|$. Each token in the vocabulary is assigned a unique numeric token id, while remaining tokens are assigned 0 or a hashing based identifier. Queries are lowercased, split on whitespace, and converted into a sequence of token ids. We truncate the query tokens at the 99th percentile by length. Token vectors that are smaller than the predetermined length are padded to the right.

Products have multiple attributes, like title, brand, and color, that are material to the matching process. We evaluated architectures to embed every attribute independently and concatenate them to obtain the final product representation. However, large variability in the accuracy and availability of structured data across products led to 5\% lower recall than simply concatenating the attributes. Hence, we decided to use an ordered bag of words of these attributes.

\section{Experiments}
\label{sec:expt}
In this section we describe our metrics, training procedure, and the results, including the impact of our method in production.

\subsection{Metrics}

We define two evaluation subtasks: matching and ranking. 
\begin{enumerate}
\item{Matching}: The goal of the matching task is to retrieve all relevant documents from a large corpus for a given query. In order to measure the matching performance, we first sample a set of 20K queries. We then evaluate the model's ability to recall purchased products from a sub-corpus of 1 million products for those queries. Note that the 1 million product corpus contains purchased and impressed products for every query from the evaluation period as well as additional random negatives. We tune the model hyperparameters to maximize Recall@100 and Mean Average Precision (MAP).
\item{Ranking}: The goal of this task is to order a set of documents by relevance, defined as purchase count conditioned on the query. The set of documents contains purchased and impressed products. We report standard information retrieval ranking metrics, such as Normalized Discounted Cumulative Gain (NDCG) and Mean Reciprocal Rank (MRR).
\end{enumerate}

\subsection{Results}
\begin{table*}
  \caption{Loss Function Experiments using Unigram Tokenization and Average Pooling}
  \label{tab:loss}
  \begin{tabular}{l|cccc|cc}
    \toprule
    Loss & Recall & MAP & Matching NDCG & Matching MRR & Ranking NDCG & Ranking MRR \\
    \midrule
    BCE & \textbf{0.586} & \textbf{0.486} & \textbf{0.695} & \textbf{0.473} & \textbf{0.711} & \textbf{0.954} \\
    MAE & 0.044 & 0.020 & 0.275 & 0.192 & 0.611 & 0.905 \\
    MSE & 0.238 & 0.144 & 0.490 & 0.377 & 0.680 & 0.948 \\
    \hline
    2 Part L1 & 0.485 & 0.384 & 0.694 & 0.472 & 0.742 & 0.966 \\
    3 Part L1 &  \textbf{0.691} &  \textbf{0.616} &  \textbf{0.762} &  \textbf{0.536} &  \textbf{0.760} &  \textbf{0.971} \\
    \hline
    2 Part L2 & 0.651 & 0.576 & 0.768 & 0.549 & \textbf{0.776*} & \textbf{0.973*} \\
    3 Part L2 & \textbf{0.735*} &  \textbf{0.664*} &  \textbf{0.791*} &  \textbf{0.591*} & 0.772 &  \textbf{0.973*} \\
    \bottomrule
  \end{tabular}
\end{table*}

\begin{table*}
  \caption{Token Embedding Aggregation Experiments using Unigram Tokenization}
  \label{tab:pool}
  \begin{tabular}{ll|cccc|cc}
    \toprule
    Loss & Pooling & Recall & MAP & Matching NDCG & Matching MRR & Ranking NDCG & Ranking MRR \\
    \midrule
    \multirow{3}{*}{MSE} & ave & \textbf{0.238} & \textbf{0.144} & \textbf{0.490} & \textbf{0.377} & 0.680 & 0.948 \\
     & gru & 0.105 & 0.052 & 0.431 & 0.348 & 0.700 & 0.951 \\
     & lstm & 0.102 & 0.048 & 0.404 & 0.286 & \textbf{0.697} & \textbf{0.948} \\ \midrule
    \multirow{3}{*}{3 Part L1} & ave & \textbf{0.691} & \textbf{0.616} & \textbf{0.762} & \textbf{0.536} & \textbf{0.760} & \textbf{0.971} \\
     & gru & 0.651 & 0.574 & 0.701 & 0.376 & 0.727 & 0.965 \\
     & lstm & 0.661 & 0.588 & 0.730 & 0.469 & 0.739 & 0.964 \\ \midrule
    \multirow{3}{*}{3 Part L2} & \textbf{ave} & 0.735 & 0.664 & \textbf{0.791*} & \textbf{0.591*} & 0.772 & 0.973 \\
     & gru & \textbf{0.739*} & 0.659 & 0.777 & 0.578 & \textbf{0.775*} & \textbf{0.975*} \\
     & lstm & 0.738 & \textbf{0.666*} & 0.767 & 0.527 & \textbf{0.775*} & \textbf{0.976*} \\
    \bottomrule
  \end{tabular}
\end{table*}

\begin{table*}
  \caption{Tokenization Experiments with Average Pooling and 3 Part L2 Hinge Loss}
  \label{tab:token}
  \begin{tabular}{l|cccc|cc}
    \toprule
    Tokenization & Recall & MAP & Matching NDCG & Matching MRR & Ranking NDCG & Ranking MRR \\
    \midrule
    Char Trigrams & 0.673 & 0.586 & 0.718 & 0.502 & 0.741 & 0.955 \\ 
    Unigrams & 0.735 & 0.664 & 0.791 & 0.591 & 0.772 & 0.973 \\

    Unigrams+Bigrams & 0.758 & 0.696 & 0.784 & 0.577 & 0.768 & 0.974 \\
    Unigrams+Bigrams+Char Trigrams & \textbf{0.764} & \textbf{0.707} & \textbf{0.800} & \textbf{0.615} & \textbf{0.794*} & \textbf{0.978} \\
    \hline
    Unigrams+OOV & 0.752 & 0.694 & 0.799 & 0.633 & 0.791 & 0.978 \\
    Unigrams+Bigrams+OOV & 0.789 & 0.741 & 0.790 & 0.610 & 0.776 & 0.979 \\
    Unigrams+Bigrams+Char Trigrams+OOV & \textbf{0.794*} & \textbf{0.745*} & \textbf{0.810*} & \textbf{0.659*} & \textbf{0.794*} & \textbf{0.980*} \\ \midrule
    Unigrams(500K) & 0.745 & 0.683 & 0.799 & \textbf{0.629} & 0.784 & 0.975 \\
    Word Unigrams(125K)+OOV(375K) & \textbf{0.753} & \textbf{0.694} & \textbf{0.804} & 0.612 & \textbf{0.788} & \textbf{0.979} \\
    \bottomrule
  \end{tabular}
\end{table*}

In this section, we present the durable learnings from thousands of experiments. We fix the embedding dimension to 256, weight matrix initialization to Xavier initialization \cite{Glorot10understandingthe}, batch size to 8192, and the optimizer to ADAM with the configuration $\alpha=0.001, \beta_1=0.9, \beta_2=0.999, \epsilon=10^{-8}$ for all the results presented. We refer to the hinge losses defined in Section \ref{sec:loss_fn} with $m=1$ and $m=2$ as the L1 and L2 variants respectively. Unigram tokenization is used in Table \ref{tab:loss} and Table \ref{tab:pool}, as the relative ordering of results does not change with other more sophisticated tokenizations. 

We present the results of different loss functions in Table \ref{tab:loss}. We see that the L2 variant of each loss consistently outperforms the L1. We hypothesize that L2 variants are robust to outliers in cosine similarity. The 3-part hinge loss outperforms the 2-part hinge loss in matching metrics in all experiments although the two loss functions have similar ranking performance. By considering impressed negatives, whose text is usually more similar to positives than negatives, separately from random negatives in the 3-part hinge loss, the scores for positives and random negatives become better separated, as shown in Section \ref{sec:loss_fn}. The model can better differentiate between positives and random negatives, improving Recall and MAP. Because the ranking task is not distinguishing between relevant and random products but instead focuses on ordering purchased and impressed products, it is not surprising that the 2-part and 3-part loss functions have similar performance. 

In Table \ref{tab:pool}, we present the results of using LSTM, GRU, and averaging to aggregate the token embeddings. Averaging performs similar to or slightly better than recurrent units with significantly less training time. As mentioned in Section \ref{sec:arch}, in the product search setting, queries and product titles tend to be relatively short, so averaging is sufficient to capture the short-range dependencies that exist in queries and product titles. Furthermore, recurrent methods are more expressive but introduce specialization between the query and title. Consequently, local word-level matching between the query and the product title may not be not captured as well.

In Table \ref{tab:token}, we compare the performance of using different tokenization methods. We use average pooling and the 3-part L2 hinge loss. For each tokenization method, we select the top $k$ terms by frequency in the training data. Unless otherwise noted, $k$ was set to 125K, 25K, 64K, and 500K for unigrams, bigrams, character trigrams, and out-of-vocabulary (OOV) bins respectively. It is worth noting that using only character trigrams, which was an essential component of DSSM\cite{huang2013dssm}, has competitive ranking but not matching performance compared to unigrams. Adding bigrams improves matching performance as bigrams capture short phrase-level information that is not captured by averaging unigrams. For example, the unigrams for ``chocolate milk'' and ``milk chocolate'' are the same although these are different products. Additionally including character trigrams improves the performance further as character trigrams provide generalization and robustness to spelling errors. 

Adding OOV hashing improves the matching performance as it allows better generalization to infrequent or unseen terms, with the caveat that it introduces additional parameters. To differentiate between the impact of additional parameters and OOV hashing, the last two rows in Table \ref{tab:token} compare 500K unigrams to 125K unigrams and 375K OOV bins. These models have the same number of parameters, but the model with OOV hashing performs better.

In Table \ref{tab:norm}, we present the results of using batch normalization, layer normalization, or neither on the aggregated query and product embeddings. The ``Query Sorted'' column refers to whether all positive, impressed, and random negative examples for a single query appear together or are shuffled throughout the data. The best matching performance is achieved using batch normalization and shuffled data. Using sorted data has a significantly negative impact on performance when using batch normalization but not when using layer normalization. Possibly, the batch estimates of mean and variance are highly biased when using sorted data. 

Finally, in Table \ref{tab:baselines}, we compare the results of our model to four baselines: DSSM \cite{huang2013dssm}, Match Pyramid \cite{pang2016text}, ARC-II \cite{Hu:2014:CNN:2969033.2969055}, and our model with frozen, randomly initialized embeddings. We only use word unigrams or character trigrams in our model, as it is not immediately clear how to extend the bag-of-tokens approach to methods that incorporate ordering. We compare the performance of using the 3-part L2 hinge loss to the original loss presented for each model. Across all baselines, matching performance of the model improves using the 3-part L2 hinge loss. ARC-II and Match Pyramid ranking performance is similar or lower when using the 3-part loss. Ranking performance improves for DSSM, possibly because the original approach uses only random negatives to approximate the softmax normalization. More complex models, like Match Pyramid and ARC-II, had significantly lower matching and ranking performance while taking significantly longer to train and evaluate. These models are also much harder to tune and tend to overfit.

The embeddings in our model are trained end-to-end. Previous experiments using other methods, including Glove and word2vec, to initialize the embeddings yielded poorer results than end-to-end training. When comparing our model with randomly initialized to one with trained embeddings, we see that end-to-end training results in over a 3X improvement in Recall@100 and MAP.

\begin{table*}
  \caption{Normalization Layer Experiments}
  \label{tab:norm}
  \begin{tabular}{ll|cccc|cc}
    \toprule
    Query Sorted & Normalization & Recall & MAP & Matching NDCG & Matching MRR & Ranking NDCG & Ranking MRR \\
    \midrule
     \multirow{3}{*}{T} & batch & 0.730 & 0.663 & 0.763 & 0.553 & 0.751 & 0.970 \\
      & layer &  \textbf{0.782} &  \textbf{0.733} &  \textbf{0.817*} &  \textbf{0.649} &  \textbf{0.812*} &  \textbf{0.982*} \\
      & none & 0.780 & 0.722 & 0.798 & 0.616 & 0.799 & 0.976 \\ \hline
     \multirow{3}{*}{F} & batch & \textbf{0.794*} & \textbf{0.745*} & \textbf{0.810} &  \textbf{0.659*} & 0.794 & \textbf{0.980} \\
      & layer & 0.791 & 0.743 & 0.807 & 0.629 &0.797 & \textbf{0.980} \\
      & none & 0.784 & 0.728 & 0.803 & 0.639 &  \textbf{0.803} & 0.976 \\
    \bottomrule
  \end{tabular}
\end{table*}

\newcolumntype{L}{>{\centering\arraybackslash}m{1.6cm}}
\begin{table*}
  \caption{Comparison with Baselines}
  \label{tab:baselines}
  \begin{threeparttable}
  \begin{tabular}{l | ll | ccLL | LL}
    \toprule
    Model & Loss & Tokenization & Recall & MAP & Matching NDCG & Matching MRR & Ranking NDCG & Ranking MRR \\ \midrule
    \multirow{2}{*}{Our Model} & 3 Part L2 & Char Trigrams & 0.673 & 0.586 & 0.718 & 0.502 & 0.741 & 0.955 \\
    & 3 Part L2 & Unigrams & \textbf{0.735} & \textbf{0.664} & \textbf{0.791} & \textbf{0.591} & \textbf{0.772} & \textbf{0.973}  \\ \midrule
    \multirow{2}{1.9cm}{Our Model (Random Emb)} & 3 Part L2 & Char Trigrams & 0.268 & 0.149 & 0.291 & 0.075 & 0.426 & 0.792 \\
    & 3 Part L2 & Unigrams &  0.207 & 0.107 & 0.249 & 0.052 & 0.412 & 0.778 \\ \midrule
    \multirow{4}{*}{DSSM \cite{huang2013dssm}} & Crossentropy$^\ddagger$ & Char Trigrams$^\ddagger$ & 0.647 & 0.537 & 0.576 & 0.278 & 0.589 & 0.903  \\
    & 3 Part L2 & Char Trigrams & 0.662 & 0.568 & 0.726 & 0.557 & 0.745 & 0.956 \\ \cline{2-9}
    & Crossentropy & Unigrams & 0.702 & 0.580 & 0.526 & 0.206 & 0.534 & 0.890 \\
    & 3 Part L2 & Unigrams & 0.702 & 0.614 & 0.704 & 0.492 & 0.738 & 0.960 \\ \midrule
    \multirow{2}{1.9cm}{Match Pyramid \cite{pang2016text}} & BCE$^\ddagger$ & Unigrams$^\ddagger$ & 0.475 & 0.357 & 0.599 & 0.348 & 0.682 & 0.959 \\
    & 3 Part L2 & Unigrams & 0.562 & 0.450 & 0.611 & 0.358 & 0.654 & 0.956  \\ \midrule 
    \multirow{2}{*}{ARC II \cite{Hu:2014:CNN:2969033.2969055}} & Pairwise$^\ddagger$ & Unigrams$^\ddagger$ & 0.399 & 0.270 & 0.547 & 0.299 & 0.673 & 0.939  \\
    & 3 Part L2 & Unigrams & 0.465 & 0.348 & 0.577 & 0.353 & 0.671 & 0.936 \\
    \bottomrule
  \end{tabular}
  \begin{tablenotes}
\item[$\ddagger$] These are the results from the best model trained using the loss and tokenization methodology presented in the original paper.
\end{tablenotes}
\end{threeparttable}
\end{table*}

\subsection{Online Experiments}
We ran a total of three online match set augmentation experiments on a large e-commerce website across three categories: toys and games, kitchen, and pets. In all experiments, the conversion rate, revenue, and other key performance indicators (KPIs) statistically significantly increased. One challenge we faced with our semantic search solution was weeding out irrelevant results to meet the precision bar for production search quality. To boost the precision of the final results, we added guard rails through additional heuristics and ranking models to filter irrelevant products. A qualitative analysis of the augmented search results coupled with an increase in relevant business metrics provided us with compelling evidence that this approach contributed to our goal of helping customers effortlessly complete their shopping missions.

	\begin{table}
		\caption{Example Queries and Matched Products}
		\label{tab:qual}
			
		\begin{tabular}{l l}
			\toprule
			\textbf{Query:} & make it bake it suncatchers \\
			\textbf{Comments:} &  Robustness to Spelling Error	\\
			\multicolumn{2}{l}{ \includegraphics[width=0.8\columnwidth]{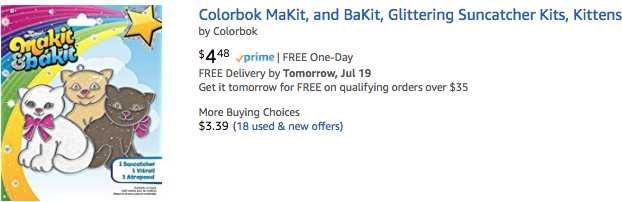}} \\
			\hline
			\textbf{Query:} & healthy shampoo  \\
			\textbf{Comments:}& Associates sulfate-free to healthy \\
			\multicolumn{2}{l}{ \includegraphics[width=0.8\columnwidth]{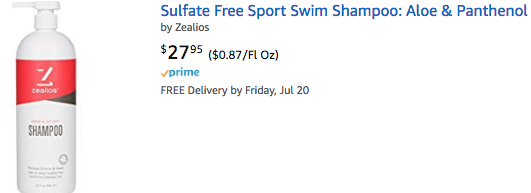}} \\
			\hline
			\textbf{Query:}	& collapsible step ladder  \\
			\textbf{Comments:} & Synonymous intent \\
			\multicolumn{2}{l}{\includegraphics[width=0.8\columnwidth]{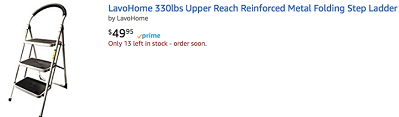}} \\
			\hline
			\textbf{Query:}	& ninjago lego training kai minifigure  \\
			\textbf{Comments:} & Drops uninformative token "training" \\
			\multicolumn{2}{l}{\includegraphics[width=0.8\columnwidth]{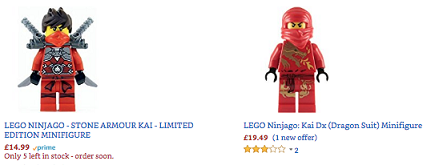}} \\
			\bottomrule	
	\end{tabular}
\end{table}

\section{Training Acceleration}
\label{sec:train_acc}
During our offline experiments, we saw an average of 10\% improvement in matching metrics by increasing the data from 200 million to 1.2 billion query-product pairs. In this section, we describe our multi-GPU training techniques for efficiently handling these larger datasets. Most parameters for this model lie in the embedding layer and hence data parallel training approaches have high communication overhead. Furthermore data parallel training limits the embedding matrix size as the model must fit in a single GPU. The simplicity of averaging pooling combined with the separability of the Siamese architecture allow us to use model parallel training to increase the throughput. Letting $k$ represent the embedding dimension and $n$ represent the number of GPUs, the similarity function of our model is shown in equation \ref{eqn:network}. The embedding matrix is split among the GPUs along the embedding dimension. The input is sent to all GPUs to look up the partial token embeddings and average them. Sending the input to all GPUs tends to be inexpensive as the number of tokens is small in comparison with the token embeddings or the embedding matrix. Simply concatenating the partial average embeddings across GPUs requires $O(2k)$ communication of floating point numbers per example in both forward and backward propagation. Equation \ref{eqn:partdot} and \ref{eqn:partsum} show how instead the cosine similarity can be decomposed to transmit only the partial-sums and partial-sum-of-squares. With this decomposition, we incur a constant communication cost of 3 scalars per GPU. 

\begin{equation}
\text{Sim}(Q,P) = \cos(E^Q, E^P)
\label{eqn:network}
\end{equation}

\begin{align}
\cos(a, b) &= \frac {a \cdot b}{\left \| a \right \|_2 \left \| b \right \|_2} = \frac{\sum\limits_{i=1}^k a_i\cdot b_i}{\sqrt{\sum\limits_{i=1}^k a_i^2}\sqrt{\sum\limits_{i=1}^k b_i^2}}\\
\intertext{Splitting the cosine similarity computation across $n$ GPUs:}
r &= k/n \\ 
\sum\limits_{i=1}^k a_i\cdot b_i &= \sum_{j=1}^{n}\sum\limits_{l=1}^r  a_{r(j-1) + l} \cdot b_{r(j-1)+l} \label{eqn:partdot} \\
\sum\limits_{i=1}^k a_i^2 &= \sum_{j=1}^{n}\sum\limits_{l=1}^r  a_{r(j-1) + l}^2 \label{eqn:partsum}
\end{align}

\begin{figure}[h]
	\centering
	\includegraphics[width=0.95\linewidth]{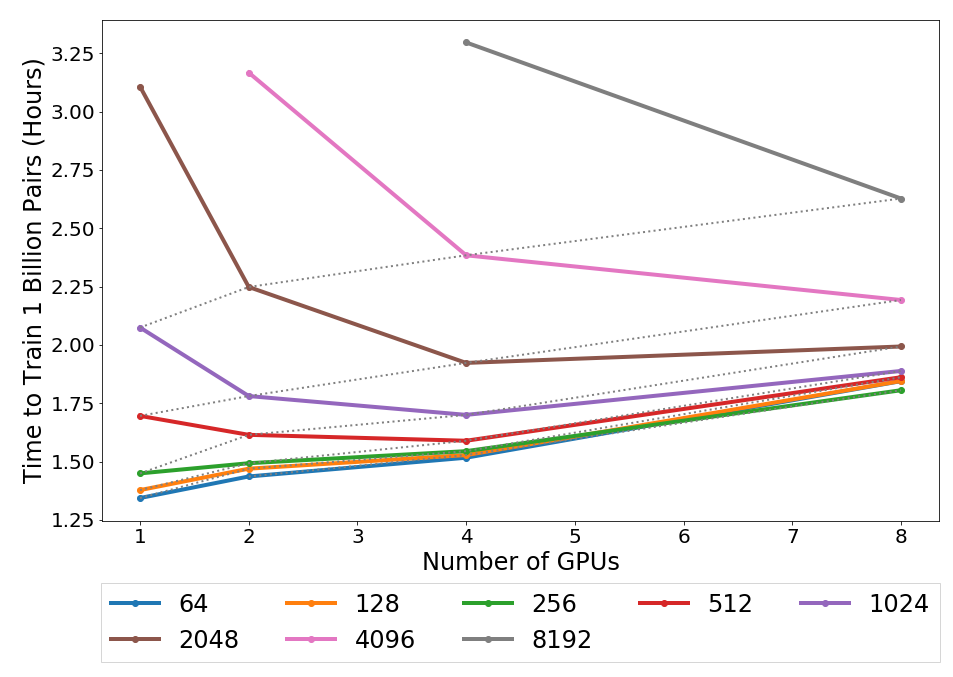}
	\caption{Training time with various embedding dimensions}
	\label{fig:multigpu}
\end{figure}

Results from these experiments are shown in Figure \ref{fig:multigpu}. We ran experiments on a single AWS p3.16xlarge machine with 8 NVIDIA Tesla V100 GPUs (16GB), Intel Xeon E5-2686v4 processors, and 488GB of RAM. The training was run 5 times with 10 million examples. The median time, scaled to 1 billion examples, is reported. 

To achieve scaling, we had to ensure that the gradient variables were placed on the same GPUs as their corresponding operations. This allows greater distribution of memory usage and computation across all GPUs. Unsurprisingly, splitting the model across GPUs for smaller embedding dimensions (<256) increases the overall training time. But beyond an embedding dimension of 512, the communication overhead is less than the additional computational power. Note that the training time is almost linear at a constant embedding dimension per GPU. In other words, training with an embedding dimension of 2048 on 2 GPUs and an embedding dimension of 1024 on 1 GPU have similar speeds. In Figure \ref{fig:multigpu}, this is shown by the dotted lines connecting points with the same embedding dimension per GPU. With ideal scaling, the lines would be horizontal.

\section{Conclusion and Future Work}
\label{sec:conclusion}
We presented our semantic product search model for an online retail store to improve product discovery with significant increases in KPIs. We discussed intuitions, practical tradeoffs, and key insights learned from many iterations of experiments. We introduced a 3-part hinge loss and showed that it outperforms other variants by deftly handling impressed but not purchased products. Furthermore, we showed that hashing unseen tokens improves the precision across different tokenization strategies. We observed significant improvements to offline metrics by increasing the training data and presented our data preprocessing approach to reduce training time. Finally, we presented our approach to training models across multiple GPUs to enable learning with larger embedding sizes and reduce the training time. 
In the future, we hope to improve the precision of our models and eliminate the need for additional heuristics to filter irrelevant results online. Our initial experiments using self-attention mechanisms and positional encodings did not show improvements in precision over our existing model, which we posit further underscores the unique nature of product search versus more traditional problems in IR and NLP. We will continue exploring approaches for scaling both training and inference. 

\begin{acks}
Nan Chen, Abhinandan Patni, and Trishul Chilimbi were instrumental in gathering data, engineering our training platform, and increasing the training speed. Yesh Dattatreya and Sunny Rajagopalan helped in hyperparameter optmization. Guy Lebanon, Vishy Vishwanathan and Inderjit Dhillon served as advisors throughout the project. Scott Le Grand and Edward Kandrot provided guidance to design and implement model parallel training.
\end{acks}

%
\bibliographystyle{ACM-Reference-Format}
\bibliography{new-ir}

%
\appendix

\begin{table*}
  \caption{Shared versus Decoupled Embeddings for Query and Product}
  \label{tab:shared_emb}
  \begin{tabular}{llc|ccLL|LL}
    \toprule
    Tokenization & Loss & Shared & Recall & MAP & Matching NDCG & Matching MRR & Ranking NDCG & Ranking MRR \\
    \midrule
    \multirow{2}{*}{Unigrams} & \multirow{2}{1.9cm}{BCE} & F & 0.520 & 0.418 & 0.649 & 0.420 & 0.692 & 0.953 \\
     &  & T & \textbf{0.586} & \textbf{0.486} & \textbf{0.695} & \textbf{0.473} & \textbf{0.711} & \textbf{0.954} \\ \midrule
    \multirow{2}{*}{Unigrams} & \multirow{2}{*}{MSE} & F & 0.131 & 0.069 & 0.389 & 0.307 & \textbf{0.690} & \textbf{0.956} \\
     &  & T & \textbf{0.238} & \textbf{0.144} & \textbf{0.490} & \textbf{0.377} & 0.680 & 0.948 \\ \midrule
    \multirow{2}{*}{Unigrams} & \multirow{2}{*}{2 Part L2} & F & 0.622 & 0.553 & \textbf{0.773} & \textbf{0.581} & 0.775 & \textbf{0.974} \\
     &  & T & \textbf{0.651} & \textbf{0.576} & 0.768 & 0.549 & \textbf{0.776} & 0.973  \\ \midrule
    \multirow{2}{*}{Unigrams} & \multirow{2}{*}{3 Part L2} & F & 0.730 & 0.662 & 0.771 & \textbf{0.593} & 0.766 & 0.972 \\
     & & T & \textbf{0.735} & \textbf{0.664} & \textbf{0.791} & 0.591 & \textbf{0.772} & \textbf{0.973} \\ \midrule
    \multirow{2}{2.8cm}{Unigrams+Bigrams+ Char Trigrams+OOV} & \multirow{2}{*}{3 Part L2} & F & 0.781 & 0.739 & 0.799 & 	\textbf{0.639*} & 0.784 & 0.980 \\ 
     &  & T & \textbf{0.790*} & \textbf{0.743*} & \textbf{0.805*} & 0.625 & \textbf{0.794*} & \textbf{0.981*} \\ 
    \bottomrule
  \end{tabular}
\end{table*}

\begin{table*}
  \caption{Impact of Out-of-Vocabulary Bin Size}
  \label{tab:oov}
  \begin{tabular}{l|ccLL|LL}
    \toprule
    Tokenization & Recall & MAP & Matching NDCG & Matching MRR & Ranking NDCG & Ranking MRR \\
    \midrule
    Unigrams+Bigrams+Char Trigrams & 0.764 & 0.707 & 0.800 & 0.615 & 0.794 & 0.978 \\
    Unigrams+Bigrams+Char Trigrams+5K OOV & 0.767 & 0.711 & 0.802 & 0.617 & 0.800 & 0.979 \\
    Unigrams+Bigrams+Char Trigrams+10K OOV & 0.774 & 0.714 & 0.811 & 0.633 & 0.804 & 0.979 \\
    Unigrams+Bigrams+Char Trigrams+50K OOV & 0.777 & 0.725 & 0.810 & \textbf{0.637} & \textbf{0.807} & 0.981 \\
    Unigrams+Bigrams+Char Trigrams+100K OOV & 0.784 & 0.733 & \textbf{0.817} & 0.629 & \textbf{0.807} & \textbf{0.982} \\
    Unigrams+Bigrams+Char Trigrams+250K OOV & \textbf{0.790} & 0.740 & 0.814 & 0.623 & 0.804 & 0.980 \\
    Unigrams+Bigrams+Char Trigrams+500K OOV & \textbf{0.790} & \textbf{0.743} & 0.805 & 0.625 & 0.794 & 0.981 \\
    \bottomrule
  \end{tabular}
\end{table*}

\section{Additional Experiments}
This section details additional experiments completed to determine the model architecture and to tune model hyperparameters. 

We demonstrate empirically in Table \ref{tab:shared_emb} that sharing the embedding layer between the query and product arm tends to perform better for matching results across multiple tokenizations and loss functions. As we described previously, sharing the embedding layer helps local word-level matching and generalization to unseen tokens when using OOV bins. Note that in this experiment, the number of model parameters was held constant. So the embedding dimension was 256 for the shared embedding layer but 128 for each of the decoupled query and product embedding layers. 

In Table \ref{tab:oov}, we present the results of varying the OOV bin size. We see that matching performance improves as the bin size increases, although ranking performance peaks at lower bin sizes. These results confirm the intuition that adding OOV hashing leads to generalization to unseen tokens. This generalization improves matching performance as there are fewer spurious matches resulting from OOV tokens mapping to the same bucket and/or simply excluding OOV tokens.

\end{document}